\documentclass[a4paper,10pt]{article}
\usepackage{geometry}
\usepackage{physics}
\usepackage{authblk}
\usepackage{graphicx}
\usepackage{caption}
\usepackage{subcaption}
\usepackage{fullminipage}
\usepackage{tocloft}
\usepackage{amsmath,amssymb}
\usepackage{slashed}
\usepackage[linkcolor=blue,citecolor=blue,colorlinks=true]{hyperref}
\usepackage{cleveref}
\crefformat{equation}{#2Eq.~(#1)#3}
\Crefformat{equation}{#2Eq.~(#1)#3}
\let\eqref\cref
\usepackage{amssymb}
\usepackage[numbers,sort&compress]{natbib}

\title{Moduli stabilization with bulk scalar in nested doubly warped braneworld model}
\author[1]{Arko Bhaumik \thanks{arkobhaumik12@gmail.com}}
\author[2]{Soumitra SenGupta \thanks{tpssg@iacs.res.in}}
\affil[1]{Physics and Applied Mathematics Unit \protect\\ Indian Statistical Institute, Kolkata-700108, India \vspace*{2mm}}
\affil[2]{School of Physical Sciences \protect\\ Indian Association for the Cultivation of Science, Kolkata-700032, India}
\date{}                     
\begin{document}

\maketitle

\begin{abstract}
We examine  the modulus stabilization mechanism of a warped geometry  model with nested warping. Such a model with multiple moduli is known to offer a possible resolution of the  fermion mass hierarchy problem in the standard model.  A six dimensional  doubly warped braneworld model under consideration admits two distinct moduli, with the associated warp factors dynamically generating different physical mass scales on four 3-branes. In order to address the hierarchy problem related to the Higgs mass, both moduli need to be stabilized around their desired values without any extreme fine tuning of parameters. We show that it is possible to stabilize them simultaneously due to the appearence of an  effective 4D moduli potential, which is generated by a single bulk scalar field having non-zero VEVs frozen on the 3-branes. We also discuss how the entire mechanism can possibly be understood from a purely gravitational point of view, with higher curvature $f(R)$ contributions in the bulk automatically providing a scalar degree of freedom that can serve as the stabilizing field in the Einstein frame.
\end{abstract}

\newpage

\tableofcontents

\section{Introduction}
The genesis of extra dimensional theories in physics dates back to the early 1920s, marked by Kaluza and Klein's original attempt to unify classical electrodynamics and Einstein gravity. In recent decades, higher-dimensional braneworld models have attracted fresh interest \cite{1990, stringref1,stringref2,add1,add2,add3, brane, braneworld, rs1, rs2,3brane,ued1,ued2,firstever}, arising as potential candidates for addressing the gauge hierarchy problem which is inherently linked to the problem of the Higgs mass. As the only fundamental scalar in the Standard Model (SM), the Higgs should experience large radiative corrections to its mass due to its self-coupling and couplings to other matter and gauge fields. These corrections are expected to flow up to the ultraviolet cut-off scale of the underlying quantum field theory. Protecting the Higgs mass from such Planck-scale corrections and keeping it safely within the TeV scale requires an extremely precise fine tuning of parameters, which leads to a naturalness problem within the SM. In this sense, the discovery of a Higgs boson as light as $125\:\textrm{GeV}/c^2$ \cite{higgs1,higgs2,higgs3} has simultaneously validated the last major prediction of the SM and exposed its central conundrum. 

The warped braneworld model proposed by Randall and Sundrum \cite{rs1} offers an elegant explanation of the large mass hierarchy, based on the premise of a non-factorizable warped five-dimensional spacetime geometry. The extra dimension is taken to be an orbifold with topology $S^1/\mathbb{Z}_2$. Two 3-branes are located at the opposite boundaries of the bulk, which is a slice of $\textrm{AdS}_5$. One of these branes (the ``TeV brane") is identified with our visible universe, on which all the SM fields are assumed to be confined to. Gravity alone is assumed to permeate through the bulk. This places the apparent scale of gravity $(M_{Pl})$ on the visible brane very close to the fundamental scale $(M_5)$. In contrast, the energy density in the bulk induces an exponential warping along the extra dimension, which causes the physical mass $m_{H}^{(ph)}$ of the Higgs on the TeV brane to be suppressed from its fundamental mass $m_{H}\sim O(M_5)$ according to $m_{H}^{(ph)}=m_{H}e^{-\pi kr_c}$, where $k\sim M_5$ and $r_c$ is the brane separation (which plays the role of the extra dimensional modulus). For $kr_c\sim11.54$, the electroweak scale is dynamically generated from the fundamental scale. But the original model contains no mechanism to stabilize the magnitude of $kr_c$ around its desired value. This problem was addressed by Goldberger and Wise \cite{goldbergerwise}, who showed that $kr_c$ can be stabilized appropriately by introducing a massive scalar field $\phi$ in the bulk with quartic interactions localized on the branes. Integrating out the extra dimensions results in an effective potential $V_{eff}(r_c)$ which stabilizes the modulus without any excessive fine tuning. This mechanism further allows one to interpret the modulus $r_c$ as the vacuum expectation value (VEV) of a dynamical radion field $\rho(x^\mu)$, i.e., $r_c=\left\langle\rho\right\rangle$, with the effective potential $V_{eff}(\rho)$ driving $\rho$ to settle at its minimum \cite{phenrad}. The mass of the radion and its coupling to matter fields on the visible brane are suggested to be $\mathcal{O}(\textrm{TeV})$, rendering it detectable, in principle, at present generation colliders. Notably, the radion is expected to be lighter than the lowest-lying KK excitations of generic bulk fields, which are typically above $\mathcal{O}(\textrm{TeV})$. This should make the radion the lightest detectable signature of the warped higher dimensional world. 

There exists a wealth of work on the dynamics of various bulk fields and KK gravitons in warped spacetime, alongside a variety of generalizations with their own phenomenological and cosmological features (for a small body of examples, see \cite{5dgauge,5dscalgrav,5dferm1,5dgrav,5dferm2,5dphen,5dexp,5dkin,5dcurved,5dcurferm,5dcosm1,5dcosm2,5dcosm3,5dcosm4,5dcosm5,5dcosm6,5dcosm7}). The original Goldberger-Wise analysis has also been extended along several avenues, eg. by assuming finite values of the brane coupling constants \cite{gwexact}, incorporating back reaction of the bulk scalar on the metric \cite{gwback1,gwback2,gwback3}, and dynamical stabilization schemes in cosmological backdrops \cite{gwdyn1,gwdyn2,gwdyn3}. However, in spite of its theoretical appeal, the original 5D RS model has been increasingly challenged by experimental data in recent years. The mass of the first excitation of the graviton KK tower in the 5D setting \cite{5dgrav} is suggested to be $\mathcal{O}(\textrm{TeV})$. Moreover, the graviton KK modes are expected to couple to visible brane matter with amplified strength, as the lowest-lying excitations cluster close to the TeV brane. These considerations subsequently inspired several unsuccessful attempts to detect the signatures of the graviton modes through various channels at the LHC \cite{gravexp1,gravexp2,gravexp3,gravexp4,gravexp5,gravexp6}. The absence of any such evidence till date has imposed serious constraints on the parameter space of the model. In particular, a small hierarchy $m_H/M_5\sim10^{-2}$ is required to explain the null results, indicating the potential appearance of new physics at least two orders of magnitude below the fundamental Planck scale. This is particularly problematic since this hierarchy requires $r_c^{-1}$ (which plays the role of the cut-off above which new physics is expected to appear) to be lowered by two orders of magnitude as well. But the sensitive dependence of the exponential warp factor on $r_c$ severely restricts this possibility, thereby limiting the efficacy of the five-dimensional setup in addressing the gauge hierarchy problem.

\section{Review of doubly warped braneworld model}
Several higher dimensional extensions of the original RS model have been proposed \cite{6d1,6d2,6d3,6d4}, with most of them introducing additional orbifolds with topology $S^1/\mathbb{Z}_2$ besides the first one. Some of these models, generalized to six dimensions, have interesting cosmological features stemming from dynamical stabilization of the extra spacelike dimensions \cite{6d5}. A particularly interesting extension \cite{doubwarp}, capable of addressing the aforementioned issues, emerges in the form of a doubly warped braneworld with topology $\left[\mathcal{M}(1,3)\times S^1/\mathbb{Z}_2\right]\times S^1/\mathbb{Z}_2$. It results in a ``brane-box" configuration, with four 4-branes forming the ``walls" of the box and four 3-branes located at the ``vertices" where adjacent 4-branes intersect. The bulk is a slice of $\textrm{AdS}_6$ with cosmological constant $\Lambda_6\sim -\tilde M^6$, where $\tilde M$ is the fundamental Planck scale. The complete bulk-brane action $(S)$, comprised of the bulk Einstein-Hilbert action $(S_6)$ and the brane tension terms $(S_5)$, is 
$$S=S_6+S_5$$
$$S_6=\int d^4xdydz\sqrt{-g_6}\left(\dfrac{1}{2}\tilde M^4R_6-\Lambda_6\right)$$
\begin{equation}
\begin{split}
S_5= & \int d^4xdydz\sqrt{-g_5}\left[V_1(z)\delta(y)+V_2(z)\delta(y-\pi)\right] \\
& + \int d^4xdydz\sqrt{-\bar g_5}\left[V_3(y)\delta(z)+V_4(y)\delta(z-\pi)\right]
\end{split}
\end{equation}
where $R_6$ is the six dimensional Ricci scalar, and $y$ and $z$ are angular coordinates charting the extra dimensions. For the current purpose, the visible 3-brane has been assumed to be devoid of any matter field, which would have otherwise introduced a further ($S_4$) term as the contribution of the matter Lagrangian. In a multiply warped setting, the brane tensions can, in general, be functions of the bulk coordinates. This is a departure from the 5D case, where the presence of only one extra dimension rules out any such dependence at the very outset. Using an RS-like ansatz to solve the resulting Einstein equations, one obtains the doubly warped metric
\begin{equation} \label{metric}
ds^2=\dfrac{\textrm{cosh}^2(kz)}{\textrm{cosh}^2(k\pi)}\left(e^{-2c|y|}\eta_{\mu\nu}dx^\mu dx^\nu+R_y^2dy^2\right)+r_z^2dz^2
\end{equation}
where $R_y$ and $r_z$ are the radii of the orbifolds, and $c$ and $k$ are constants interconnected through
\begin{equation} \label{relation}
c=\dfrac{R_yk}{r_z\textrm{cosh}(k\pi)}\qquad;\qquad k=r_z\sqrt{-\dfrac{\Lambda_6}{10\tilde M^4}}
\end{equation}
This peculiar relation makes it clear that in absence of a considerably large hierarchy between $R_y$ and $r_z$, both $c$ and $k$ cannot be simultaneously large. Instead, any one of the two distinct regimes $c>k$ or $c<<k$ can emerge. It is not entirely possible to get rid of a little hierarchy between $R_y$ and $r_z$, but with suitable choice of $c$ and $k$, it can be ensured that $R_y/r_z$ does not exceed $\mathcal{O}(10)$. The physical mass scale $m_H^{(ph)}$ on the 3-brane located at the vertex $(y_i,z_j)$ is related to the fundamental scale $m_H$ by
\begin{equation} \label{massord}
m_H^{(ph)}=m_He^{-cy_i}\dfrac{\textrm{cosh}(kz_j)}{\textrm{cosh}(k\pi)}
\end{equation}
In the $c>k$ regime, the choice of $c\sim10$ and $k\sim0.1$ yields warp factors which together generate the TeV scale on the two 3-branes at $\{y=\pi,z=0\}$ and $\{y=\pi,z=\pi\}$, one of which can be identified with the visible brane. The other couple of 3-branes experiences negligible warping and remains at the Planck scale. This clustering of multiple 3-branes around each of the scales is a salient property of generalized RS models with nested warping, one having important phenomenological implications. 

Once the metric is fully solved, the brane tensions can be derived using appropriate junction conditions at the orbifold fixed points; $V_1$ and $V_2$ retain their $z$-dependence, whereas $V_3$ and $V_4$ turn out to be independent of $y$. 
\begin{equation}
V_1(z)=-V_2(z)=8\tilde M^2\sqrt{-\dfrac{\Lambda_6}{10}}\textrm{sech}(kz)
\end{equation}
\begin{equation}
V_3(y)=0\qquad;\qquad V_4(y)=-8\tilde M^2\sqrt{-\dfrac{\Lambda_6}{10}}\textrm{sech}(k\pi)
\end{equation}
The coordinate dependence of the former pair can be shown to emerge from suitable scalar fields confined to the corresponding 4-branes. In the large $k$ regime, one needs to account for a phantom scalar field on the $y=\pi$ brane. This field may serve as a potential candidate for dark energy on the TeV branes. But the presence of such a phantom scalar renders the setup an effective field theory as opposed to a fundamental one. No such phantom field is needed in the large $c$ regime. 

The doubly warped model has numerous phenomenological advantages over its progenitor model. Firstly, owing to the presence of two extra dimensions, the first excited graviton KK mode turns out to be considerably heavier than that of the 5D case. Moreover, coupling between graviton KK modes and SM fields on the visible brane is largely suppressed compared to the 5D model. Taken together, these features can satisfactorily explain the non-detection of KK gravitons at the LHC so far, without recourse to any small hierarchy between $m_H$ and $\tilde M$ \cite{doubgrav}. At the same time, significant portions of the parameter space of the extended model remain accessible to the LHC, allowing them to be explored in future runs \cite{doubgrav2}. Secondly, the doubly warped model can offer an explanation of the mass hierarchy among the SM fermions \cite{doubwarp}. In the $c>k$ regime, the $\mathcal{O}$(\textrm{TeV}) 4-brane at $y=\pi$ intersects two other 4-branes at $z=0$ and $z=\pi$. Assuming SM fermions to be described by five dimensional fields confined to the $y=\pi$ brane implies natural $\mathcal{O}$(\textrm{TeV}) masses for the fermions. In addition to the $z$-dependent bulk wavefunction on this 4-brane, the fermionic fields can have kinetic terms on the 3-branes at the two intersection points. These boundary terms can modify the fermion-scalar Yukawa coupling on the $\{y=\pi,z=0\}$ and $\{y=\pi,z=\pi\}$ 3-branes, thereby causing a splitting among the effective fermion masses. This splitting is arguably small, as the natural mass scales of these 3-branes are already clustered close to each other around the TeV scale.

\section{Modulus stabilization in doubly warped model}

Like the 5D RS model, the action of the 6D scenario contains no dynamics which can stabilize the extra dimensional moduli around their desired values. In absence of such an underlying stabilizing mechanism, the braneworld model alone cannot be considered adequate. Motivated by the success of the 5D Goldberger-Wise mechanism, it becomes natural to seek a similar approach for the 6D model that might stabilize both $c$ and $k$ (or equivalently, $R_y$ and $r_z$) simultaneously. While other phenomenological aspects of multiply warped spacetimes are well-studied \cite{higgsgauge,multscal,gaugematt1,gaugematt2}, there has been little work in this direction so far, with the notable exception of \cite{ArunEtAl}. The latter study proposes disjoint stabilization mechanisms for $c$ and $k$ with the help of two separate bulk and brane-localized fields in the $c>k$ regime. In the other regime, taking $c<<k$ makes the metric almost conformally flat, wherefore only $r_z$ needs to be stabilized satisfactorily, with $R_y$ either left unstabilized (which is justified because of the negligibly small value of $c$) or stabilized with the help of another brane-localized field. Whether stabilization of both moduli can be achieved with the help of a single bulk scalar field, has, however, been an open question so far. In this paper, we attempt to address this very question.

\subsection{Dynamics of the bulk field}

For our current purpose, we choose to work in the $c>k$ regime, which, as noted already, obviates the need to introduce a phantom scalar (which raises phenomenological difficulties) on the $y=\pi$ brane to explain the coordinate dependence of its tension. Additionally, the relative smallness of $k$ allows us to view the entire setup as a not-too-large departure from the 5D model, helping identify the key points of deviation from the latter clearly. A study of the complementary $c<<k$ domain using tools from supersymmetric quantum mechanics appears in \cite{ArunEtAl}.

Analogous to the original Goldberger-Wise scenario, we consider a bulk field propagating freely through the extra dimensional $y-z$ bulk, and interacting only at the locations of the four 3-branes through quartic self-interaction terms. The action $(S_{GW})$ of the bulk field is given by an immediate generalization of the original 5D Goldberger-Wise action.
\begin{equation} \label{action}
S_{GW}=-\dfrac{1}{2}\int d^4x\int\limits_{-\pi}^{+\pi}dy\int\limits_{-\pi}^{+\pi}dz \sqrt{-g_6}\left(g_6^{AB}\partial_A\phi\partial_B\phi+m^2\phi^2\right)-\sum_{i=1}^4\int d^4x\sqrt{\tilde g_4^{(i)}}\lambda_i\left(\phi^2-u_i^2\right)^2
\end{equation}
Here, $\tilde{g}_4^{(i)}$ is the induced metric on the $i^{th}$ 3-brane, with $u_i$ being the VEV of the bulk field (with mass dimension $[u_i]=+2$) and $\lambda_i$ the corresponding coupling constant ($[\lambda_i]=-4$). With its dynamics confined to the higher dimensional space, the bulk field is essentially ``frozen" on the corner branes. Defining $a(y)=\textrm{exp}(-c|y|)$ and $b(z)=\textrm{sech}(k\pi)\textrm{cosh}(kz)$, we extremize this action with respect to $\phi$ to obtain the following equation of motion.
\begin{equation} \label{eqmot}
\begin{split}
& -\dfrac{1}{R_y^2}\partial_y\left(a^4b^3\partial_y\phi\right)-\dfrac{1}{r_z^2}\partial_z\left(a^4b^5\partial_z\phi\right)+m^2a^4b^5\phi \\
& +\dfrac{4a^4b^4}{R_yr_z}\left[\lambda_1\phi\left(\phi^2-u_1^2\right)\delta(y)\delta(z)+\lambda_2\phi\left(\phi^2-u_2^2\right)\delta(y-\pi)\delta(z)\right] \\
& +\dfrac{4a^4b^4}{R_yr_z}\left[\lambda_3\phi\left(\phi^2-u_3^2\right)\delta(y)\delta(z-\pi)+\lambda_4\phi\left(\phi^2-u_4^2\right)\delta(y-\pi)\delta(z-\pi)\right]=0
\end{split}
\end{equation}
Away from the boundaries, the contributions of the interaction terms vanish. Assuming a separable solution of the form $\phi(y,z)=\phi_1(y)\phi_2(z)$, the bulk equation of motion can be reduced to the following pair of uncoupled ODEs, where $-4\alpha^2$ is the separation constant.
\begin{equation} \label{eq:mot}
-\dfrac{1}{R_y^2a^4\phi_1}\dfrac{d}{dy}\left(a^4\dfrac{d\phi_1}{dy}\right)=\dfrac{1}{r_z^2b^3\phi_2}\dfrac{d}{dz}\left(b^5\dfrac{d\phi_2}{dz}\right)-m^2b^2=-4\alpha^2
\end{equation}
An interesting physical interpretation of $\alpha$ emerges immediately. From the equation for $\phi_1$, it is apparent that $2\alpha$ plays the role of the mass of the $\phi_1$ field. Also, as the $k\to0$ (and simultaneously $r_z\to0$) limit implies $b(z)\to1$ and mathematically reduces the 6D metric to the familiar singly warped form, it must also enforce $2\alpha\to m$ in order to reduce this equation to the 5D Goldberger-Wise bulk equation of motion. As for the equation for $\phi_2$, it is satisfied trivially in this limit, as $\phi_2(z)\to1$ leads merely to $m^2=4\alpha^2$. However, from a physical standpoint, one cannot of course allow $k$ to be arbitrarily small, as any effective field theory in the semiclassical approach strictly remains valid only for $r_z\geq M^{-1}$, with quantum gravity effects dominating for $r_z< M^{-1}$. In this sense, the $k\to0$ limit is not a physically tenable one, but comes with a lower cut-off regulated by $\Lambda_6$. The bottom line is simply that for sufficiently small $k$, one can expect $\alpha$ to be reasonably close to $m$, which is a fact that proves useful in due course.

\eqref{eq:mot} can be solved exactly for the component fields $\phi_1(y)$ and $\phi_2(z)$, leading to the following general solution for the bulk field under the condition of $\mathbb{Z}_2$ orbifold symmetry. 
\\
\begin{equation} \label{sol}
\phi(y,z)=e^{2c|y|}\left(Ae^{\nu c|y|}+Be^{-\nu c|y|}\right)\left[DP_n^l(\textrm{tanh}(k|z|))+EQ_n^l(\textrm{tanh}(k|z|))\right]\textrm{sech}^\frac{5}{2}(kz)
\end{equation}
\\
where $A$, $B$, $D$ and $E$ are four arbitrary constants (arising on account of each equation being a second order ODE), and $P_n^l$ and $Q_n^l$ are associated Legendre functions of the first and second kind respectively. The quantities $\nu$, $n$ and $l$ are defined as
\begin{equation} \label{par1}
\nu=2\sqrt{1+\dfrac{R_y^2\alpha^2}{c^2}}
\end{equation}
\begin{equation} \label{par2}
n=2\sqrt{1+\dfrac{\alpha^2r_z^2\textrm{cosh}^2(k\pi)}{k^2}}-\dfrac{1}{2}
\end{equation}
\begin{equation} \label{par3}
l=\dfrac{1}{2}\sqrt{25+\dfrac{4m^2r_z^2}{k^2}}
\end{equation}
\\
Assuming the magnitudes of the brane coupling constants to be large (i.e. $\lambda_i\to\infty$) allows the identification of simple energetically favourable configurations to serve as boundary conditions. The very structure of the interaction terms, given by $\lambda_i(\phi^2-u_i^2)^2\delta(y-y_0)\delta(z-z_0)$, necessitates $\phi(y_0,z_0)\to u_i$ as such a configuration on the corner brane at $(y_0,z_0)$. This approach leads to the following four equations, where the shorthand $\tau_z=\textrm{tanh}(k|z|)$ has been introduced for convenience.
\\
\begin{equation} \label{bound1}
\phi_1(0)\phi_2(0)=(A+B)\left[DP_n^l(0)+EQ_n^l(0)\right]\approx u_1 
\end{equation}
\begin{equation} 
\phi_1(\pi)\phi_2(0)=e^{2c\pi}(Ae^{\nu c\pi}+Be^{-\nu c\pi})\left[DP_n^l(0)+EQ_n^l(0)\right]\approx u_2
\end{equation}
\begin{equation} 
\phi_1(0)\phi_2(\pi)=(A+B)\left[DP_n^l(\tau_\pi)+EQ_n^l(\tau_\pi)\right]\textrm{sech}^\frac{5}{2}(k\pi)\approx u_3
\end{equation}
\begin{equation} \label{bound4}
\phi_1(\pi)\phi_2(\pi)=e^{2c\pi}(Ae^{\nu c\pi}+Be^{-\nu c\pi})\left[DP_n^l(\tau_\pi)+EQ_n^l(\tau_\pi)\right]\textrm{sech}^\frac{5}{2}(k\pi)\approx u_4
\end{equation}
\\
\eqref{bound1}$-$\eqref{bound4} are not all linearly independent, and as such, only allow three of the constants to be solved in terms of the remaining fourth. This poses no problem though, as the forms of the solutions still allow $\phi(y,z)$ to be determined uniquely. In the large $c$ regime, the normalized solutions of the individual component fields are 
\begin{equation} \label{sol1}
\phi_1(y)=\sqrt{u_1}e^{2c|y|}(c_1e^{\nu c|y|}+c_2e^{-\nu c|y|})
\end{equation}
\begin{equation} \label{sol2}
\phi_2(z)=\sqrt{u_1}\left[c_3P_n^l(\tau_z)+c_4Q_n^l(\tau_z)\right]\textrm{sech}^\frac{5}{2}(kz)
\end{equation}
\\
where $c_2=1$, and the other three constants are given by 
\\
\begin{equation} \label{c1}
c_1=e^{-(\nu+2)c\pi}\left[\dfrac{u_2}{u_1}-e^{-(\nu-2)c\pi}\right]
\end{equation}
\begin{equation} \label{c3}
c_3=\dfrac{Q_n^l(\tau_\pi)-\left(\frac{u_4}{u_2}\right)Q_n^l(0)\textrm{cosh}^\frac{5}{2}(k\pi)}{P_n^l(0)Q_n^l(\tau_\pi)-Q_n^l(0)P_n^l(\tau_\pi)}
\end{equation}
\begin{equation} \label{c4}
c_4=\dfrac{\left(\frac{u_4}{u_2}\right)P_n^l(0)\textrm{cosh}^\frac{5}{2}(k\pi)-P_n^l(\tau_\pi)}{P_n^l(0)Q_n^l(\tau_\pi)-Q_n^l(0)P_n^l(\tau_\pi)}
\end{equation}
\\
It is interesting to note that, apart from the overall normalization factor of $u_1$, the dynamics of $\phi(y,z)$ is controlled not by the absolute values of the VEVs but only by their ratios. This feature is reminiscent of the 5D mechanism. Further, \eqref{sol1}$-$\eqref{c4} dictate $u_4/u_2=u_3/u_1$. So the two ratios appearing explicitly in the solution are sufficient to completely specify all the six possible ratios among the four VEVs, thus ruling out any ambiguity. 

\subsection{The effective potential}

Having obtained the solution of the bulk field, one needs to substitute it back in equation \eqref{action} and integrate over $y$ and $z$ in order to obtain the stabilizing potential $V_{eff}(c,k)$, or equivalently, $V_{eff}(R_y,r_z)$. From the resultant effective 4D action $S_{eff}$, the definition $S_{eff}=-\int d^4xV_{eff}(c,k)$ allows the potential to be read off directly. In the large $\lambda_i$ limit, this amounts to evaluating only the bulk contribution. First, let us define the following dimensionless parameters.
\begin{equation}
\mu_1=\dfrac{mr_z}{k}\quad;\quad\mu_2=\dfrac{\alpha r_z}{k}
\end{equation}
These quantities roughly estimate the ratios between the mass parameters of the bulk field and the fundamental Planck scale. From the semiclassical standpoint, the admissible range of each ratio should be $0<\mu_i<1$. In terms of these ratios, the parameters from \eqref{par1}$-$\eqref{par3} can be re-expressed as $\nu=2\sqrt{1+\mu_2^2\textrm{cosh}^2(k\pi)}$, $n=\nu-0.5$, and $l=5/2\sqrt{1+4\mu_1^2/25}$. Upon substituting the solution from \eqref{sol} in the action, the integral over $y$ can be readily evaluated, but the presence of the special functions makes the $z$-integral analytically intractable. Defining the dimensionless potential $\tilde{V}_{eff}=V_{eff}/u_1^2$, and making use of \eqref{relation} to eliminate $R_y$ and $r_z$ in favour of $c$ and $k$, the following form emerges for the potential
\\
\begin{equation} \label{pot}
\begin{split}
\tilde{V}_{eff}(c,k)=&\left[\dfrac{c_1^2(\nu+2)^2}{2\nu}e^{2\nu c\pi}+\dfrac{c_2^2(\nu-2)^2}{2\nu}-2\pi c_1c_2(\nu^2-4)c\right]\times F_1(k) \\
& +\left(\dfrac{c_1^2}{2\nu}e^{2\nu c\pi}+\dfrac{c_2^2}{2\nu}+2\pi c_1c_2c\right)\times F_2(k)
\end{split}
\end{equation}
\\
where the functions $F_1(k)$ and $F_2(k)$, arising out of the $z$-integral, are given explicitly by
\begin{equation} \label{f1}
F_1(k)=k\textrm{sech}^4(k\pi)\int\limits_{-\pi}^{+\pi}dz\left(c_3P_n^l(\tau_z)+c_4Q_n^l(\tau_z)\right)^2\textrm{sech}^2(kz)
\end{equation}
\begin{footnotesize}
\begin{equation} \label{f2}
\begin{split}
F_2(k)=\dfrac{1}{4}k\textrm{sech}^4(k\pi)\int\limits_{-\pi}^{+\pi}dz & \left[2(l-n-1)(c_3P_{n+1}^l(\tau_z)+c_4Q_{n+1}^l(\tau_z))+(2n-3)\tau_z(c_3P_n^l(\tau_z)+c_4Q_n^l(\tau_z))\right]^2 \\
& +k\textrm{sech}^4(k\pi)\int\limits_{-\pi}^{+\pi}dz\:\mu_1^2(c_3P_n^l(\tau_z)+c_4Q_n^l(\tau_z))^2
\end{split}
\end{equation}
\end{footnotesize}
There is little choice but to evaluate $F_1(k)$ and $F_2(k)$ numerically for different values of $\mu_1$, $\mu_2$, and $u_4/u_2$, which, by this point, serve as three of the fundamental parameters of the model. The fourth parameter $u_2/u_1$, contained in the coefficient $c_1$, contributes chiefly to the $c$-dependence of the potential. Equipped with the potential given by \eqref{pot}$-$\eqref{f2}, we are in a position to demonstrate the existence of a simultaneous minimum of $\tilde{V}_{eff}$ in $c$ and $k$ over some region of the parameter space that doesn't require excessive fine tuning.

\subsection{Stabilizing $k$}
Owing to the explicit form of $c_1$ from \eqref{c1}, both the linear and quadratic terms in $c_1$ appearing in the coefficients of $F_1(k)$ and $F_2(k)$ in \eqref{pot} are suppressed at least by $\mathcal{O}(e^{-4c\pi})$. For $c\sim10$, this suppression factor is nearly of order $10^{-68}$. So the leading order terms within both coefficients are the $c_2^2$ terms which suffer no such suppression. In order to study the $k$-dependence of the potential, it suffices to approximate $\tilde V_{eff}$ by taking only the dominant contributions offered by these two terms. Recalling $c_2=1$, the potential effectively reduces to a single-variable function of $k$ given by
\begin{footnotesize}
\begin{equation}
\tilde V_{eff}(c,k)\to\dfrac{1}{2}F(k)\:\approx\:\dfrac{1}{4\sqrt{1+\mu_2^2\textrm{cosh}^2(k\pi)}}\left[4\left(\sqrt{1+\mu_2^2\textrm{cosh}^2(k\pi)}-1\right)^2F_1(k)+F_2(k)\right]
\end{equation}
\end{footnotesize}
Establishing the stabilization of $k$ is thus tantamount to locating a suitable minimum of $F(k)$. The plots in figure 1 show the behaviour of $F(k)$ for various combinations of the parameters.
\\
\begin{figure}[h] 
\centering
\begin{minipage}{.3\textwidth}
\centering
\includegraphics[width=\linewidth, height=0.17\textheight]{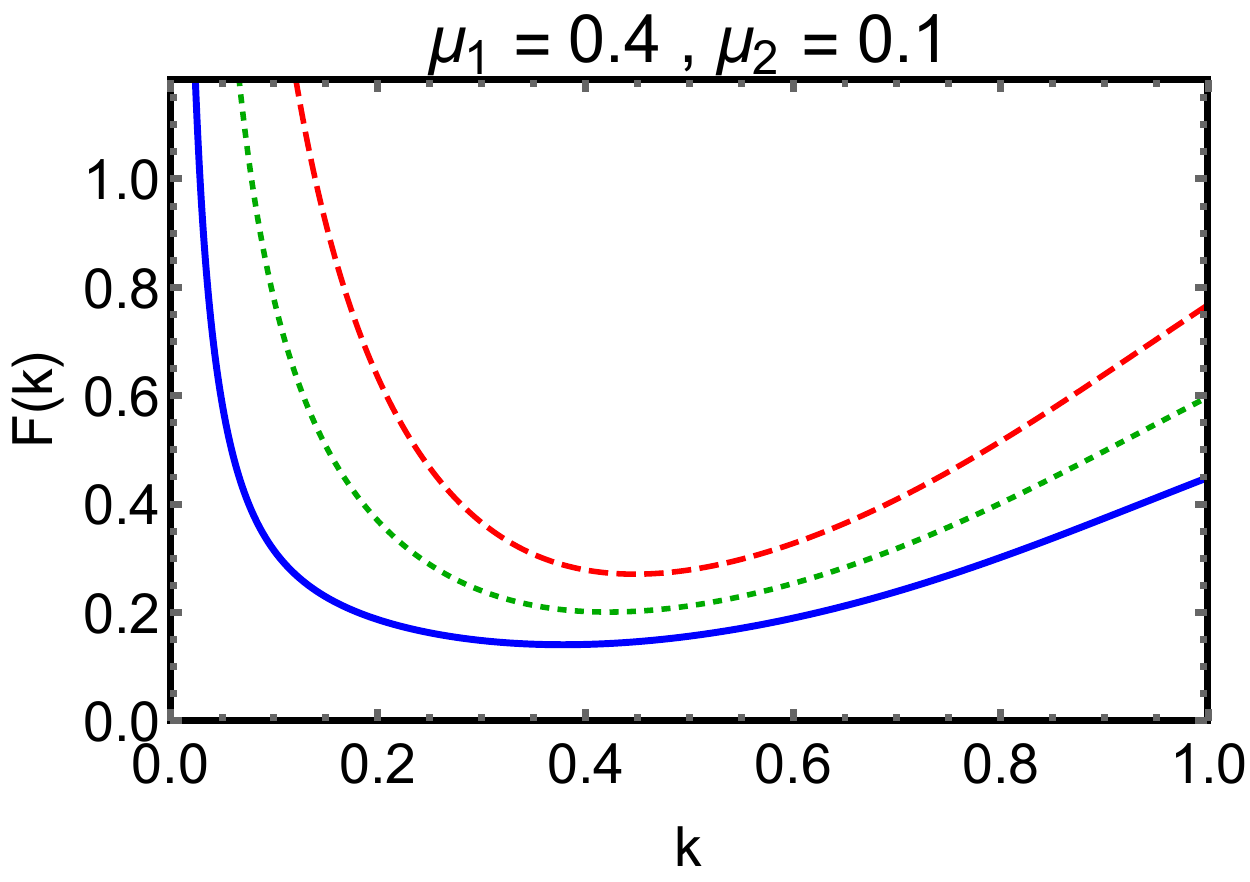}
\end{minipage}
\begin{minipage}{.3\textwidth}
\centering
\includegraphics[width=\linewidth, height=0.17\textheight]{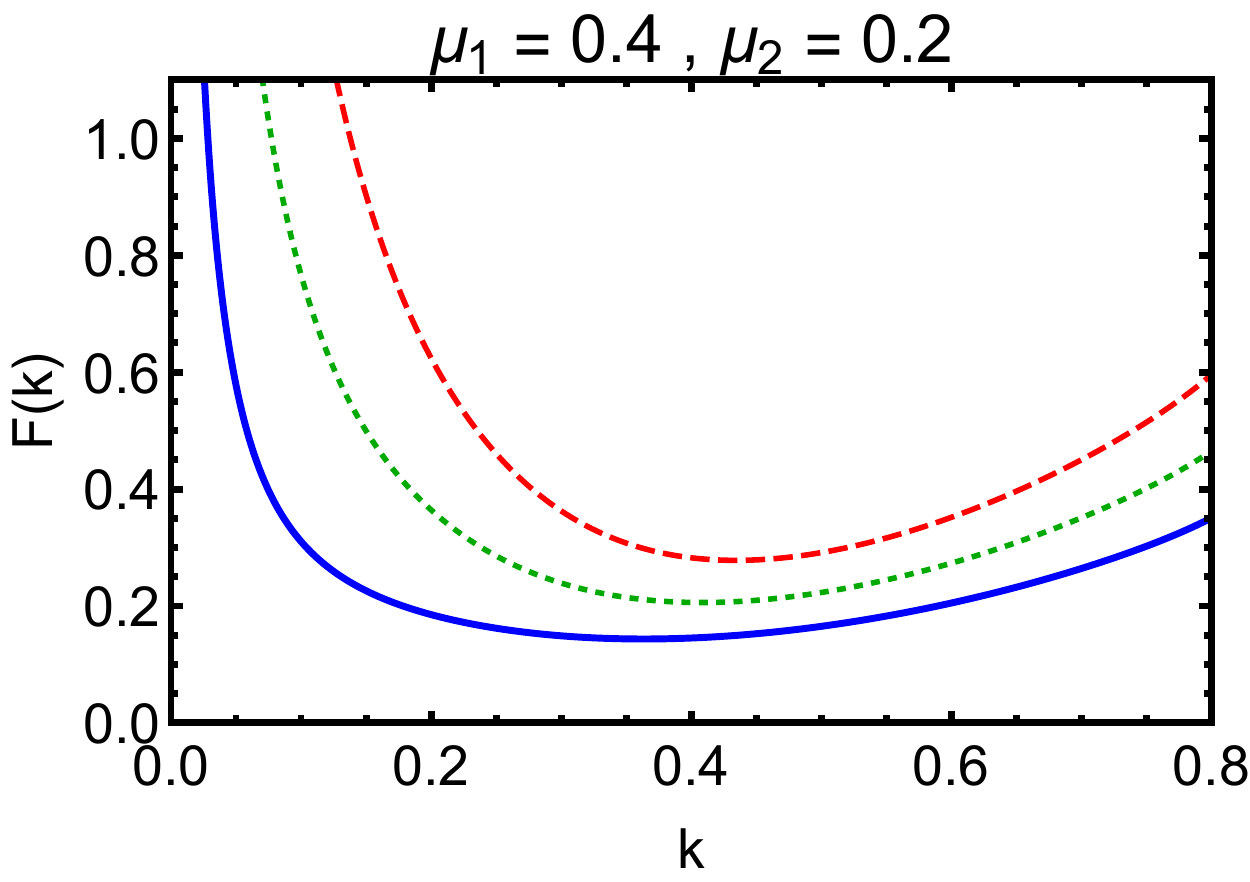}
\end{minipage}
\begin{minipage}{.3\textwidth}
\centering
\includegraphics[width=\linewidth, height=0.17\textheight]{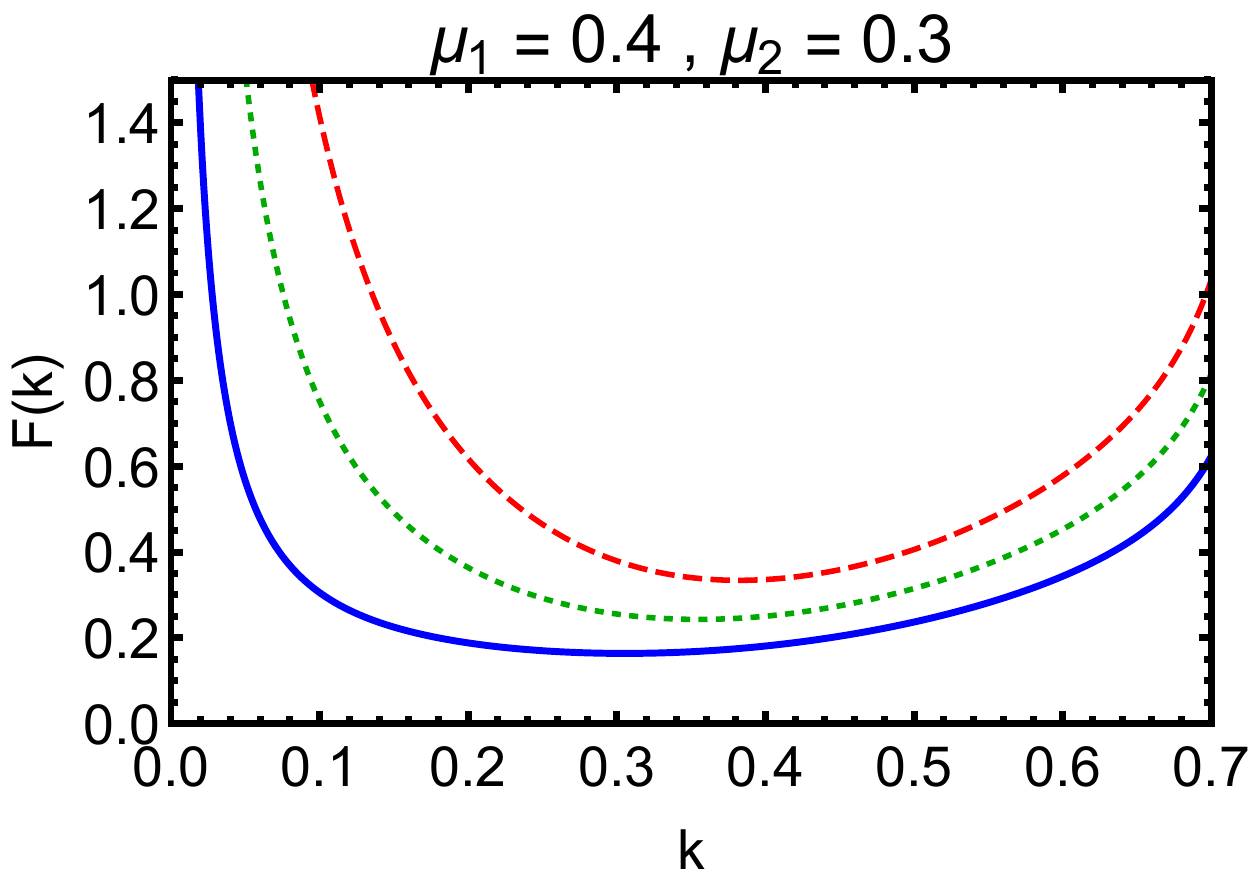}
\end{minipage}
\begin{minipage}{.3\textwidth}
\centering
\includegraphics[width=\linewidth, height=0.17\textheight]{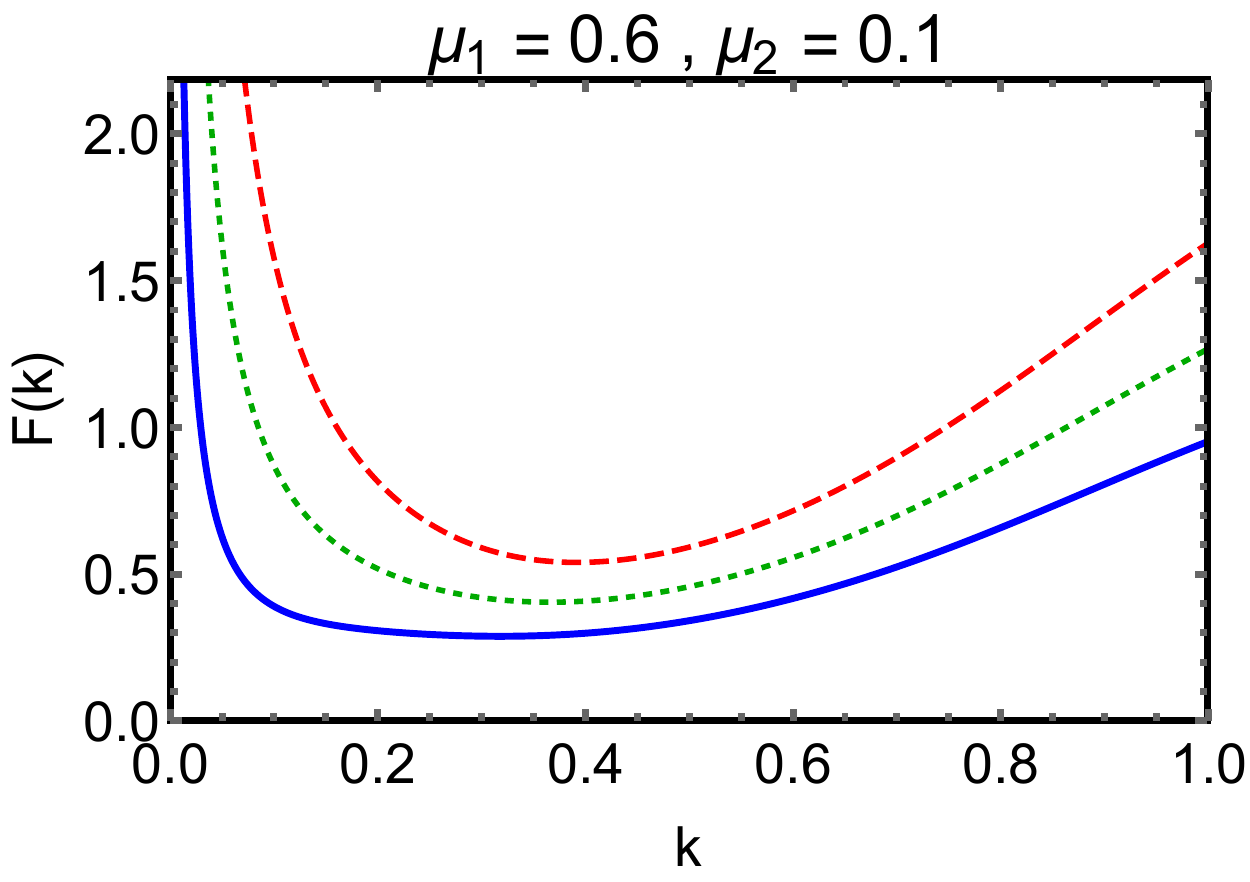}
\end{minipage}
\begin{minipage}{.3\textwidth}
\centering
\includegraphics[width=\linewidth, height=0.17\textheight]{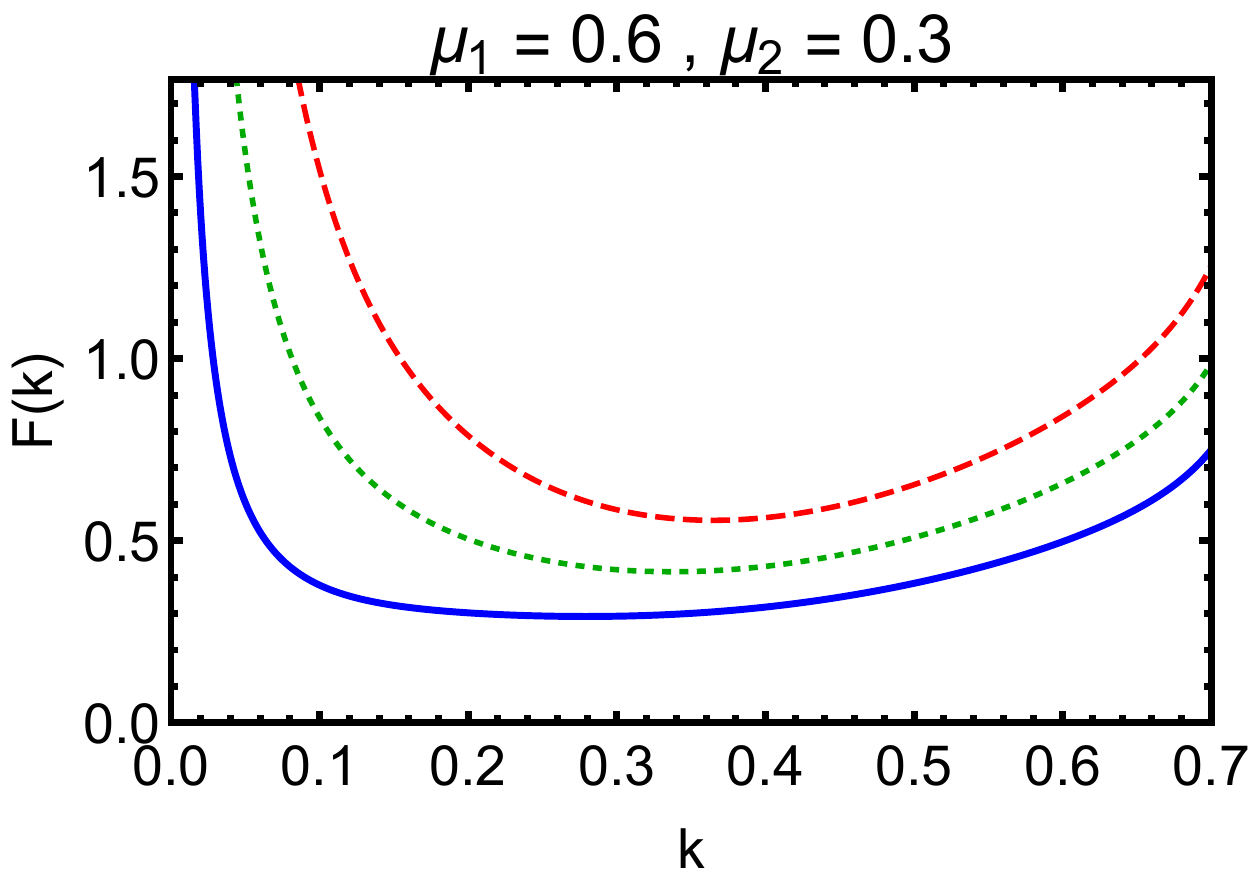}
\end{minipage}
\begin{minipage}{.3\textwidth}
\centering
\includegraphics[width=\linewidth, height=0.17\textheight]{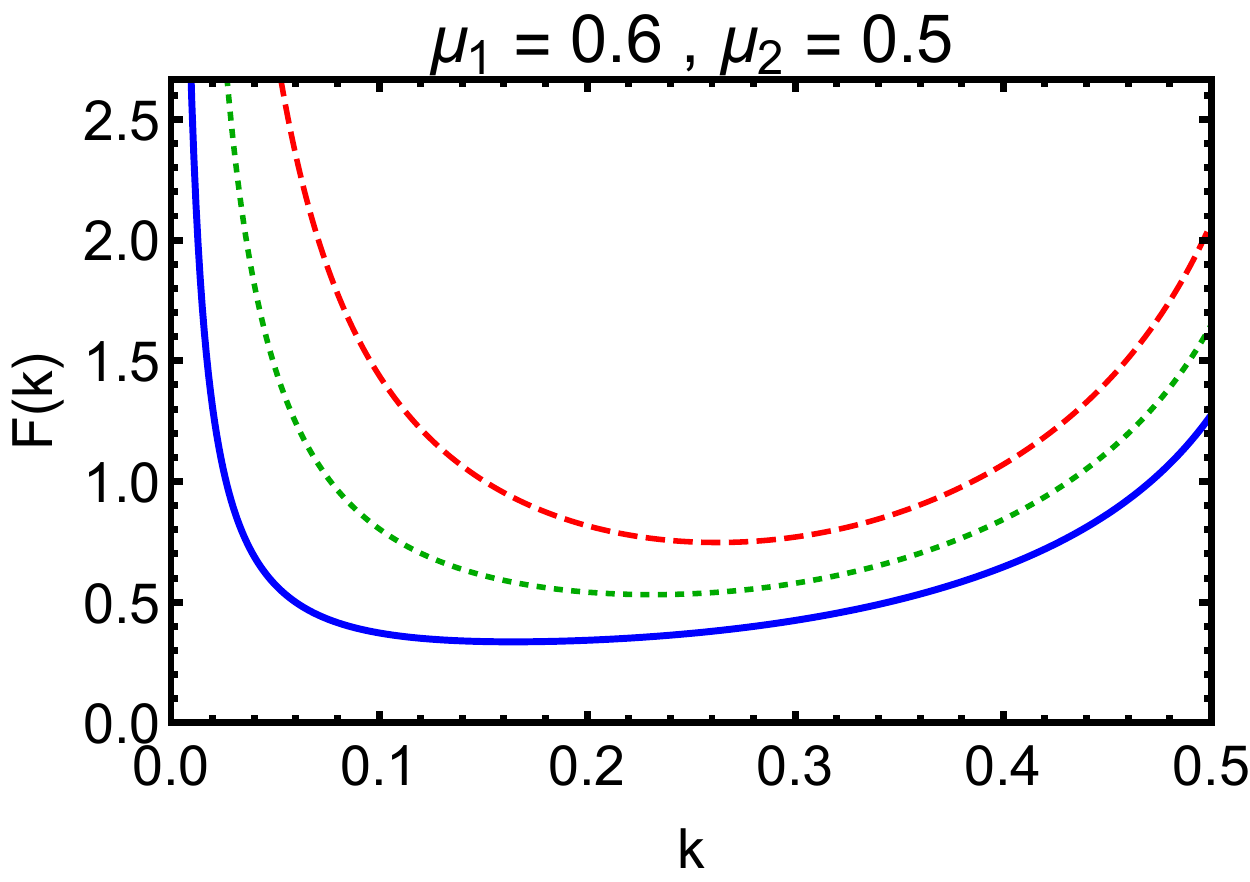}
\end{minipage}
\begin{minipage}{.3\textwidth}
\centering
\includegraphics[width=\linewidth, height=0.17\textheight]{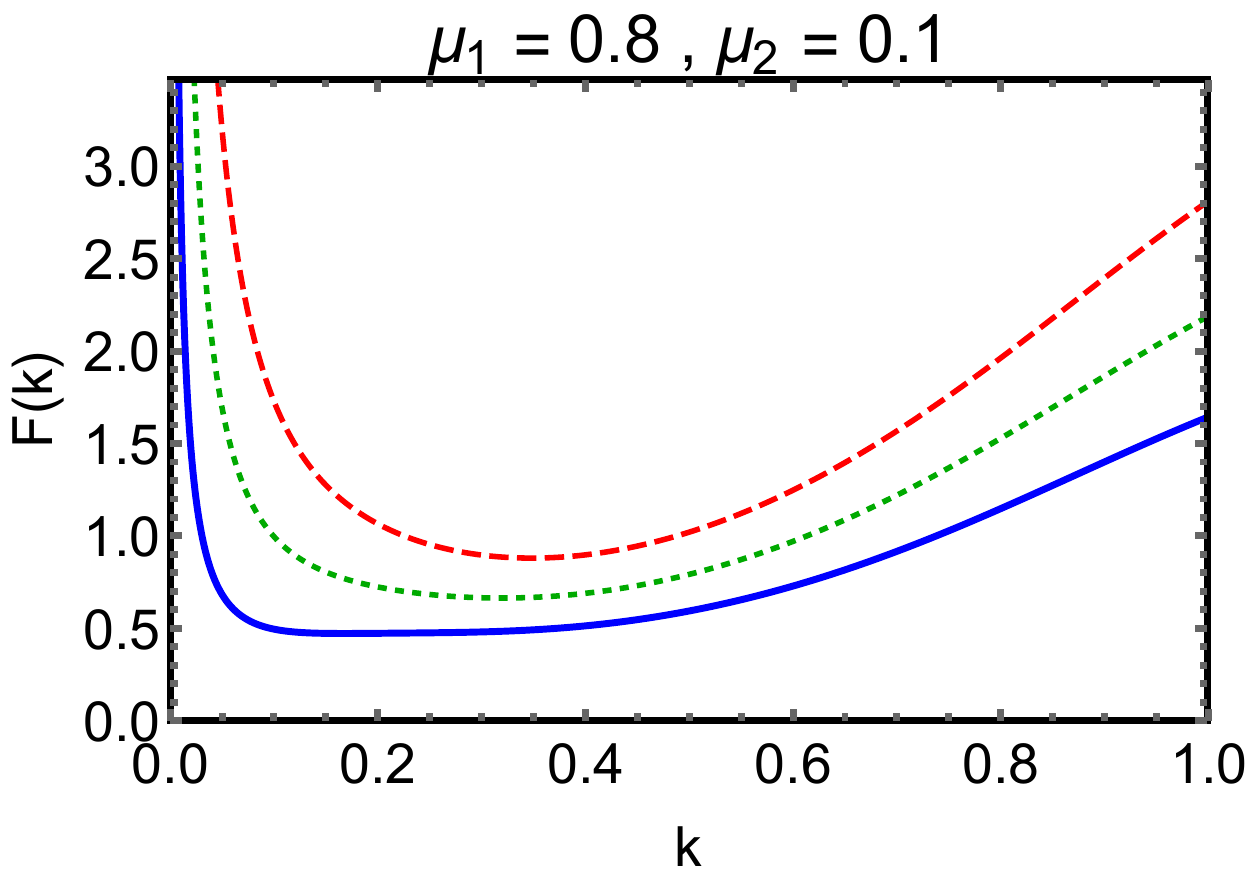}
\end{minipage}
\begin{minipage}{.3\textwidth}
\centering
\includegraphics[width=\linewidth, height=0.17\textheight]{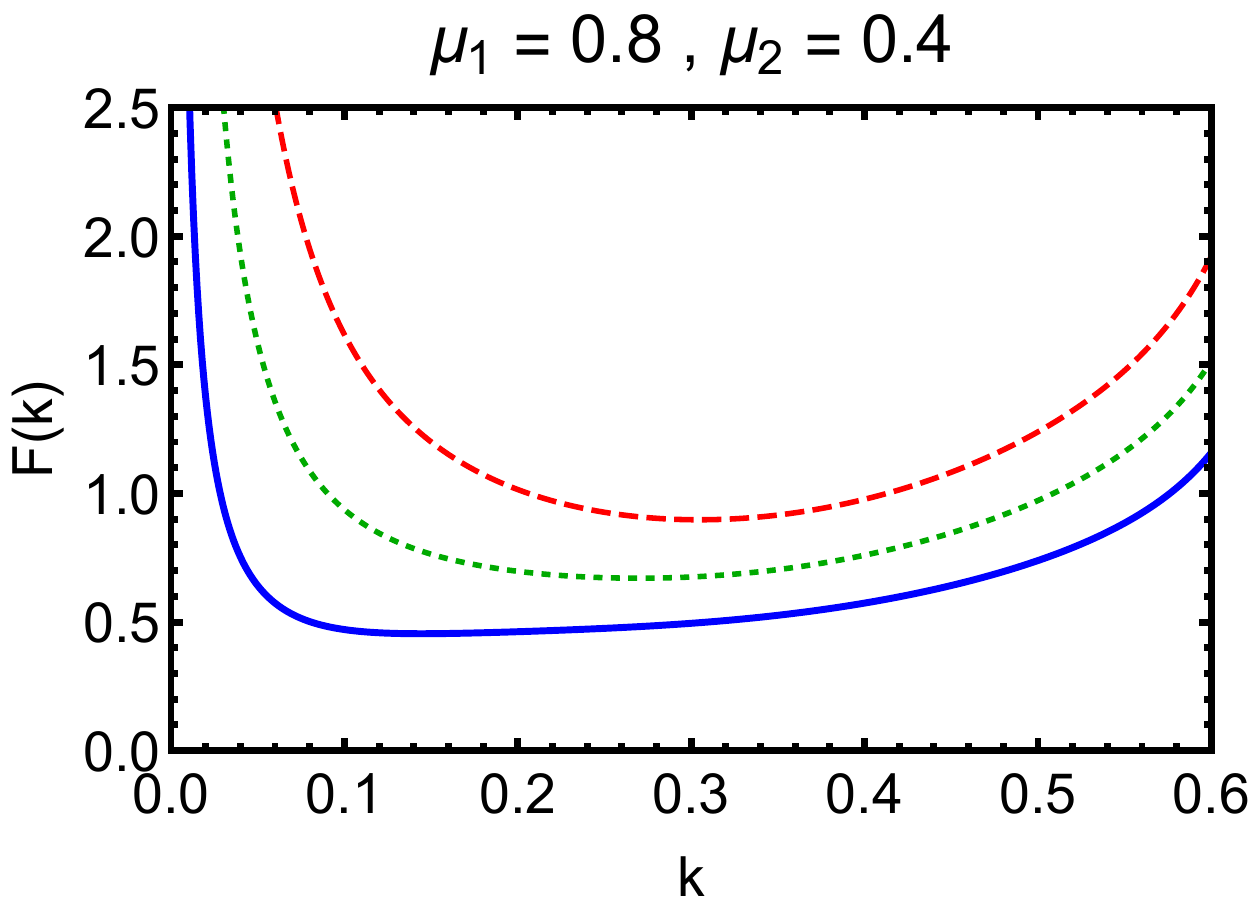}
\end{minipage}
\begin{minipage}{.3\textwidth}
\centering
\includegraphics[width=\linewidth, height=0.17\textheight]{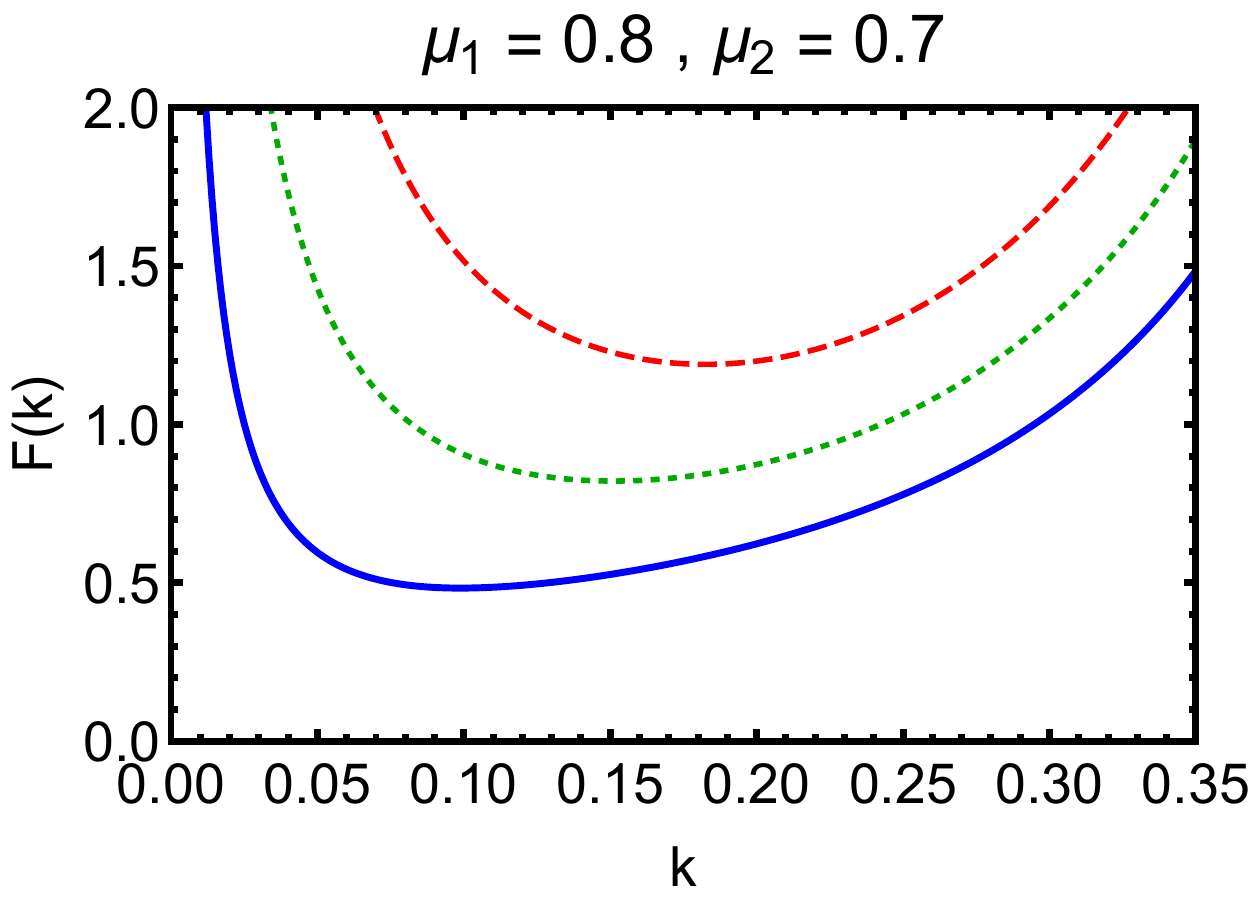}
\end{minipage}
\caption{\textit{Profiles of $F(k)$ vs $k$ for VEV ratios $u_4/u_2=1.3$ (solid blue), $u_4/u_2=1.5$ (dotted green), and $u_4/u_2=1.7$ (dashed red), and a few representative values of $\mu_1$ and $\mu_2$ for each case.}} 
\end{figure}

For all the choices of parameters, $F(k)$ admits a broad minimum at some $k<0.5$. The location of this minimum is physically important, as the range $0.1\leq k\leq0.6$ is of particular interest to scenarios attempting to explain the non-detectability of graviton KK modes at the LHC \cite{doubgrav}\cite{doubgrav2}, or exploring the phenomenology of off-brane SM fields that extend into the bulk \cite{gaugematt1}\cite{gaugematt2}. As special cases, it can be checked that $\mu_2\to0$ produces an asymptotically decaying $F(k)$ with no finite minimum, whereas $u_4=u_2$ admits only $k=0$ as the global minimum. As expected, the most pronounced dependence of the minimum is on the VEV ratio $u_4/u_2$, which governs the potential at the leading order. 

\subsection{Stabilizing $c$}

The previous plots reveal the existence of a fairly large parameter space which can stabilize $k$ around its desired value. But $\tilde{V}_{eff}(c,k)$ also needs to stabilize $c$, which plays the leading role in determining the degree of warping. To this end, we need to reinstate the previously neglected $c$-dependence in $\tilde V_{eff}(c,k)$ in order to locate a suitable minimum along $c$. Owing to the form of $c_1$ in \eqref{c1}, one might expect such a minimum to occur at $c_1=0$, which would be close in spirit to the 5D Goldberger-Wise result. Upon closer inspection though, it becomes clear that the existence of such a minimum depends crucially on the value of $\mu_2$. For considerably small values of $\mu_2$ which render $\nu\:\approx\:2+\epsilon_k$, where $\epsilon_k=\mu_2^2\textrm{cosh}^2(k\pi)$ is small enough so that $\mathcal{O}(\epsilon_k^2)$ and higher are negligible compared to unity, the deviation $\delta\tilde V_{eff}$ for any arbitrarily small deviation $c=c_0-\delta$ from the alleged minimum $c_0$ can be estimated as
\begin{equation}
\delta\tilde V_{eff}|_{c_0}\sim e^{-4\pi(c_0-\delta)-2\epsilon_kc_0\pi}\left[25\left(e^{\epsilon_k\delta\pi}-1\right)^2-8\pi(c_0-\delta)e^{\epsilon_k\delta\pi}\left(e^{\epsilon_k\delta\pi}-1\right)\left(1-6\epsilon_k\right)\right]
\end{equation}

For excessively small $\epsilon_k$ and vanishingly small $\delta$, this quantity is negative. This clearly rules out a minimum at $c_0$. The trouble can be traced to the existence of the $\mathcal{O}(c_1)$ terms in $\tilde V_{eff}(c,k)$. While the 5D stabilization mechanism ensured automatic cancellation of the pair of $\mathcal{O}(c_1)$ terms in the potential, the current model offers no such way out by default. However, if some particular combination of parameters can result in such cancellation (without excessive fine tuning of course), the issue would be resolved. Once $k$ has been stabilized at some specific $k_{min}$, the condition for the $\mathcal{O}(c_1)$ terms to cancel each other translates to
\begin{equation} \label{crit}
\left[\nu(k_{min})^2-4\right]F_1(k_{min})-F_2(k_{min})=0
\end{equation}
As $F_1(k)$ and $F_2(k)$ themselves depend on all three parameters in a rather complicated manner, this equation is transcendental and cannot be solved analytically. However, it can be checked numerically that for every choice of $\mu_1$, the allowed parameter space contains approximate solutions to \eqref{crit}, some of which are presented in Table 1. 
\\
\begin{table}[h]
\begin{center}
\begin{tabular}{|c|c|c|c|c|c|c|c|}
\hline
$\mu_1$ & $\mu_2$ & $u_4/u_2$ & $k_{min}$ & $\nu(k_{min})$ & $F_1(k_{min})$ & $F_2(k_{min})$ & $u_1/u_2$ \\
\hline \hline
0.40 & 0.24 & 1.122 & 0.119 & 2.065 & 0.697 & 0.185 & 20 \\
0.50 & 0.30 & 1.160 & 0.123 & 2.102 & 0.741 & 0.307 & 40 \\
0.60 & 0.36 & 1.200 & 0.124 & 2.145 & 0.777 & 0.469 & $\sim10^2$ \\
0.70 & 0.42 & 1.248 & 0.128 & 2.197 & 0.833 & 0.689 & $\sim10^3$ \\ 
0.80 & 0.48 & 1.303 & 0.132 & 2.256 & 0.899 & 0.980 & $\sim10^4$ \\
\hline
\end{tabular}
\caption{A few examples of critical parameter values approximately satisfying \eqref{crit} and allowing simultaneous stabilization of both moduli around $k_{min}\sim0.1$ and $c_{min}\sim10$.}
\end{center}
\end{table}

Each of the combinations ensures that the non-negative $\mathcal{O}(c_1^2)$ terms dominate over the $\mathcal{O}(c_1)$ terms by nearly two orders of magnitude, leading to a proper minimum $c_{min}$ (satisfying $c_1=0$) with the following approximate form
\begin{equation} \label{cmin}
c_{min}\:\approx\:\dfrac{1}{\pi\left[\nu(k_{min})-2\right]}\textrm{ln}\left(\dfrac{u_1}{u_2}\right)
\end{equation}
One can immediately estimate the corresponding $u_1/u_2$ ratio which produces the desired value of $c_{min}\sim11.54$. As given in Table 1, its magnitude depends strongly on $\mu_1$. In order to stabilize $k$ between $0.1$ and $0.6$, the minimum admissible magnitude of $u_1/u_2$ is $\mathcal{O}(10)$, which is slightly larger than the requirement in 5D. For larger values of $\mu_1$, a minor hierarchy between $u_1$ and $u_2$ becomes increasingly prominent. This hierarchy has its physical origin in the fact that we have essentially made the component fields $\phi_1(y)$ and $\phi_2(z)$ yield two distinct warping scales $k\sim0.1$ and $c\sim10$ \textit{in spite} of having chosen their mass parameters ($m$ and $\alpha$) to be of comparable magnitudes! The somewhat large $u_1/u_2$ ratio is precisely the price we need to pay for that. As a reality check, it is instructive to check that choosing $u_1\sim u_2$ in \eqref{cmin} would have resulted in $c_{min}\sim0.1$, just as this argument suggests. 

The mutual cancellation of the $\mathcal{O}(c_1)$ terms involves a certain degree of fine tuning among the three parameters which control $F_1(k)$ and $F_2(k)$. While the situation is clearly more delicate than the 5D case, this tuning need not be very extreme. The $\mathcal{O}(c_1^2)$ terms can dominate and provide $c$ with a stable minimum if the cancellation of the $\mathcal{O}(c_1)$ terms in \eqref{crit} is accurate roughly up to $\mathcal{O}(10^{-2})$, which is sufficient to suppress the latters' contribution by $\mathcal{O}(10^{-2})$. For $c\sim10$, the critical values of the parameters need to be accurate at most up to $\mathcal{O}(10^{-3})$, while larger values of $c$ are somewhat more likely to ameliorate the situation due to increased suppression. This is no worse than the fine tuning associated with the choice of $m/M$ for a given VEV ratio, that is required to generate the TeV scale accurately in the 5D case. Moreover, for some fixed $\nu(k)$, the tuning associated with $u_2/u_1$ can be even more lenient due to the logarithmic dependence of $c_{min}$ on $u_2/u_1$. So simultaneous stabilization of $c$ and $k$ depends on a small degree of fine tuning among the parameters $\mu_1$, $\mu_2$, and $u_4/u_2$, which constitutes a crucial aspect of the extended Goldberger-Wise mechanism in the doubly warped scenario. This feature is not surprising since one should physically expect a two-level tuning for the stabilization of two distinct moduli in a spacetime with nested warping. As the stabilization of $k$ alone requires negligible tuning (as evident from Fig. 1), the stabilization of $c$ justifiably involves both levels. It might be interesting to investigate if incorporating back reaction or quartic self-interaction terms within the bulk can lead to improvements for this tuning requirement. 

\subsection{Phenomenological implications}
The minor hierarchy between $u_1$ and $u_2$ (or equivalently between $u_3$ and $u_4$) can be better interpreted if we arrange the magnitudes of the four VEVs in proper order. As the boundary conditions require $u_1/u_2=u_3/u_4$, the VEVs obtained in each case satisfy
\begin{equation} \label{vevord}
u_3>u_1>u_4>u_2
\end{equation}
The first two values are very close to each other, as are the last two, with the aforesaid hierarchy pushing the pairs apart. Interestingly, \eqref{vevord} reflects the order of physical mass scales on the corresponding corner branes (on which these classical values of $\phi$ are defined) for large $c$ and small $k$, as can be checked using \eqref{massord} . This can be understood physically as the massive bulk scalar, once frozen on the boundary branes, naturally tends to act against the warping induced by the bulk energy density. Consequently, the resulting mass scales on corner branes associated with larger VEVs can be generally expected to be higher than those with smaller VEVs. This feature is visible in the 5D Goldberger-Wise mechanism as well, albeit in a greatly tempered form. In the present case, it is more pronounced as it also incorporates the clustering effect observed among the physical mass scales.

Due to the smallness of $k$, phenomenological features associated with $c$ should be sufficiently similar to those of the 5D model. At first glance, this might make the smallness of $u_2$ compared to $u_1$ for larger values of $\mu_1$ appear alarming, as the mass of the associated radion in the 5D model is proportional to the ratio between the classical value of $\phi$ on the TeV brane and the fundamental scale. In particular, one might worry that such small $u_2$ could bring the predicted radion mass down to the MeV scale or lower, thereby giving rise to phenomenological issues. It must be borne in mind, however, that we have neglected the back reaction of the bulk field. On occasions, this can cause underestimation of the radion mass by a few orders of magnitude \cite{gwback1}\cite{gwback2}. Furthermore, unlike the 5D model, one can exploit the 6D setup immensely by allowing SM fields to extend into the bulk. This possibility, which constitutes one of the most interesting features of the doubly warped model, might imply further corrections to the radion mass due to interactions of $\phi$ with these fields. There can also be significant mixing between $\phi$ and the scalar degrees of freedom which give rise to coordinate dependent 4-brane tensions. Any realistic study of radion phenomenology needs to address these questions in order to estimate the magnitude of the radion mass accurately. 

Besides, due to the presence of two distinct moduli in the model, one should physically expect the appearance of two fundamental radions with different masses and widths. Owing to the smallness of $k$, the second radion should not suffer any significant suppression from the fundamental scale. This is echoed by the form of the stabilizing potential $\tilde V_{eff}(c,k)$ from \eqref{pot}, where $F(k)$ is usually $\mathcal{O}(1)$ around its minimum. However, the aforementioned subtleties need to be accounted for in this case too before its phenomenology can be deduced conclusively. Due to the evidently complicated nature of such a study, we defer it to a future work.

\section{Insight from higher curvature gravity}
One of the most compelling advantages of having the size of the extra dimensions set by a single bulk field (as opposed to separate bulk and brane localized fields) is that it allows a purely gravitational interpretation of the stabilizing mechanism. Conventional wisdom suggests that the Einstein-Hilbert action, which provides an effective low energy description of gravity, needs to be amended with additional higher curvature terms respecting diffeomorphism invariance at sufficiently high energy scales. The warped geometry model which is being considered here has a large cosmological constant ( $\sim M_P$ ) in the bulk and as a result the inclusion of higher curvature terms is a natural  choice. Two broad classes of such higher curvature  theories are the quasi-linear Lanczos-Lovelock models and $f(R)$ models. While Lanczos-Lovelock models enjoy the benefit of being naturally ghost-free \cite{lanclove1,lanclove2,lanclove3}, the mathematically simpler $f(R)$ models, equipped with specific conditions to ensure freedom from ghosts, pass some of the  cosmological tests \cite{fr1,fr2,fr3,fr4,fr5,fr6}. Furthermore, any given $f(R)$ action typically admits a dual scalar-tensor representation \cite{frrev1,frrev2,frrev3,frrev4}. In the so-called Einstein frame (related to the Jordan frame through a conformal transformation), the situation is equivalent to that of a scalar field $\tilde\phi$ coupled minimally to gravity, alongside a potential $U(\tilde\phi)$ whose form is determined by the functional form of $f(R)$. For singularity-free metrics which are not experiencing rapid evolution, this equivalence holds physically \cite{physeq1,physeq2,physeq3,physeq4,physeq5}. 

In recent works \cite{gravstab1}\cite{gravstab2}, it has been shown how such a scalar degree of freedom, arising solely from gravity in the 5D RS model, can play the role of the bulk field in the Goldberger-Wise scheme, thus obviating the need to introduce the latter by hand. By choosing $f(R)=R+aR^2-|b|R^4$ (where $a$ and $b$ are coupling constants satisfying $a>0$ and $a>|b|$ to ensure freedom from ghosts), the potential can be arranged to contain both quadratic and quartic terms, with the latter encapsulating the effects of back reaction. Although the technique can be readily extended to spacetimes of arbitrary dimensionality, the inclusion of back reaction quickly renders multiply warped settings intractable in their full generality. In the following analysis, we discuss an extension of this technique to the doubly warped model, by retaining only the lowest-order correction term in $f(R)$. To that end, we choose $f(R)=R+a R^2$, which provides the action
\begin{equation} \label{fr}
\mathcal{A}=\int d^6x\sqrt{-g}\left[\dfrac{f(R)}{2\kappa_6^2}-\Lambda_6\right]=\int d^6x\sqrt{-g}\left[\dfrac{1}{2\kappa_6^2}\left(R+a R^2\right)-\Lambda_6\right]
\end{equation}
where $\kappa_6$ is the six-dimensional gravitational constant. In order to arrive at the Einstein frame, one conventionally makes a detour \cite{frrev4} through the intermediate Jordan frame representation
\begin{equation} \label{jord}
\mathcal{A}=\int d^6x\sqrt{-g}\left[\dfrac{1}{2\kappa_6^2}\{\psi R-V(\psi)\}-\Lambda_6\right]\quad,\quad V(\psi)=\chi(\psi)\psi-f(\chi(\psi))
\end{equation}
where one introduces the auxiliary field $\chi$ and defines $\psi=f'(\chi)$. For $f''(\chi)\neq0$, the equation of motion for $\chi$ from \eqref{jord}, i.e. the on-shell condition, imposes $\chi=R$. This makes \eqref{jord} equivalent to the original $f(R)$ action in \eqref{fr}. In order to reduce this to the minimally coupled Einstein frame representation, we apply the conformal transformation
\begin{equation} \label{trans}
g_{AB}\to\tilde g_{AB}=\sqrt{\psi}g_{AB}\quad,\quad\kappa_6\tilde\phi=\dfrac{\sqrt{5}}{2}\:\textrm{ln}\:\psi=\dfrac{\sqrt{5}}{2}\:\textrm{ln}\:f'(R)
\end{equation}
where the last equality follows from the on-shell condition. The action then transforms to
\begin{equation} \label{eins}
\mathcal{A}=\int d^6x\sqrt{-\tilde g}\left[\left(\dfrac{\tilde R}{2\kappa_6^2}-\Lambda_6\right)-\dfrac{1}{2}\partial_A\tilde\phi\partial^A\tilde\phi-U(\tilde\phi)\right]\quad,\quad U(\tilde\phi)=\dfrac{Rf'(R)-f(R)}{2\kappa_6^2(f'(R))^\frac{3}{2}}
\end{equation}
Substituting $f(R)=R+a R^2$ in \eqref{trans} and inverting the relation yields $R(\tilde\phi)$, which can be plugged immediately in \eqref{eins} to obtain $U(\tilde\phi)$ as follows.
\begin{equation}
R=\dfrac{1}{2a}\left(e^{\frac{2}{\sqrt{5}}\kappa_6\tilde\phi}-1\right)\quad\implies\quad U(\tilde\phi)=\dfrac{1}{8a\kappa_6^2}\left(e^{\frac{1}{\sqrt{5}}\kappa_6\tilde\phi}-2e^{-\frac{1}{\sqrt{5}}\kappa_6\tilde\phi}+e^{-\frac{3}{\sqrt{5}}\kappa_6\tilde\phi}\right)
\end{equation}
The minimum of $U(\tilde\phi)$ occurs at $\tilde\phi=0$, as evident from $U'(0)=0$ and $U''(0)>0$. Expanding $U(\tilde\phi)$ about this minimum, the leading order non-vanishing contribution comes from the quadratic term, with all subsequent terms increasingly suppressed by higher powers of $\kappa_6$.
\\
\begin{footnotesize}
\begin{equation}
U(\tilde\phi)\:\approx\:\dfrac{1}{8a\kappa_6^2}\left[\left(1+\dfrac{\kappa_6\tilde\phi}{\sqrt{5}}+\dfrac{\kappa_6^2\tilde\phi^2}{10}\right)-2\left(1-\dfrac{\kappa_6\tilde\phi}{\sqrt{5}}+\dfrac{\kappa_6^2\tilde\phi^2}{10}\right)+\left(1-\dfrac{3\kappa_6\tilde\phi}{\sqrt{5}}+\dfrac{9\kappa_6^2\tilde\phi^2}{10}\right)\right]=\dfrac{\tilde\phi^2}{10a}
\end{equation}
\end{footnotesize}
\\
Having identified $1/5a$ with the mass squared of the scalar mode $\tilde\phi$, the action from \eqref{eins} reduces precisely to the sum of the Einstein-Hilbert action and the Goldberger-Wise action in the 6D bulk, with $\tilde\phi$ in the Einstein frame remarkably playing the role of the bulk field. 

While \cite{gravstab2} uses this approach to deal with the full back-reacted problem (by including the $R^4$ term in the $f(R)$ Lagrangian) for the 5D RS model in the Jordan and Einstein frames separately, such an exact treatment is too complicated to be feasible in the 6D case. Instead, with the results from the gravitational sector at hand, one can use heuristic arguments to bridge the gap with the existing bulk field method from the preceding sections. As the physical origin of the higher curvature correction(s) can be traced to the bulk energy density, it stands to reason that the resulting scalar $\tilde\phi$ should be an explicit function of the coordinates $y$ and $z$ only. With the metric solution in the Einstein frame given by \eqref{metric}, the equation of motion for $\tilde\phi$ is identical to \eqref{eqmot} away from the boundaries. This prevents $\tilde\phi$ from having any non-trivial dynamics on the 3-branes, which is in agreement with the bulk field prescription. As an immediate corollary, we obtain four constant values of $\tilde\phi$ (having mass dimension $+2$) serving as fixed boundary values on the four branes. The results are completely analogous to \eqref{bound1}$-$\eqref{bound4}, with the only difference being that the current approach renders these boundary conditions exact, whereas in the earlier formulation they were valid only in the large coupling regime. With the full solution \eqref{sol1}$-$\eqref{c4} for $\tilde\phi$ subsequently at hand, the rest of the analysis can proceed exactly as before, leading to the stabilization of both moduli under appropriate choices of parameters.

\section{Discussions}
The prospect of stabilizing both moduli of a doubly warped Randall-Sundrum braneworld model using a single bulk scalar field has been studied. Such a mechanism is crucial for a complete resolution of the gauge hierarchy problem in a higher dimensional scenario, and needs to supplement the gravitational part of the action giving rise to the warped metric. While the approach taken here is essentially a direct generalization of the Goldberger-Wise mechanism to six dimensions, the presence of nested warping brings out additional subtleties and constraints on the parameter space. As demonstrated, these constraints do not necessarily involve any extreme fine tuning of the fundamental parameters. In the $c>k$ regime, which is phenomenologically preferred as it requires no brane-localized phantom field to explain the coordinate dependence of the 4-brane tensions, the effective potential admits a true minimum in $k$ around $k_{min}\sim0.1$ without any significant tuning. The stabilization of $c$, on the other hand, requires tuning on two different levels: firstly among the parameters $\mu_1$, $\mu_2$, and $u_4/u_2$ (for the appearance of a suitable $c_{min}$), and secondly for $u_1/u_2$ (to ensure $c_{min}\sim12$). This can be interpreted physically as a consequence of the doubly warped structure of the underlying spacetime. A further departure from the singly warped model is that in order to achieve $0.1\leq k_{min}\leq0.6$ and $c_{min}\sim12$, the minimum admissible magnitude of $u_1/u_2$ is $\mathcal{O}(10)$, which is one order of magnitude higher than the analogous requirement in case of the 5D mechanism. 

The bulk scalar approach is especially attractive as it allows room for a purely gravitational interpretation, with higher curvature contributions in the bulk automatically giving rise to the required scalar mode and its potential in the Einstein frame. Since the bulk of such warped geometry model is endowed with large bulk cosmological constant , the contributions from  higher curvature terms become natural. This motivates us to include higher curvature terms in the bulk such as in $f(R)$ model.  This feature distinguishes it from certain other stabilization schemes, eg. stabilization of the two moduli with two distinct bulk and brane-localized fields. For appropriate choices of $f(R)$, a variety of bulk scalar potentials can be generated, the simplest of which is the quadratic potential considered here. While the singly warped model could be solved exactly in presence of non-negligible back reaction, the doubly warped model becomes unsolvable as the field equations turn out to be non-linear PDEs. Nevertheless, the well-known $R^2$ correction is sufficient to make contact with the conventional approach on physical grounds. In soothe, at sufficiently high energy scales, gravity alone appears capable of both warping spacetime and determining the degree of warping. This possibility is intriguing, as it may produce observable TeV scale signatures of higher curvature gravity that can be explored in future collider experiments. In principle, foremost among them should be the $\mathcal{O}(\textrm{TeV})$ radion mass associated with the larger modulus, which would be of fundamental importance in estimating/constraining the magnitude(s) of the higher curvature coupling(s). 

The current study can be extended along various avenues. As pointed out earlier, a realistic study of radion phenomenology in a multiply warped background needs to include interactions of the bulk field with higher dimensional fermionic and gauge fields into account, alongside mixing with the brane-localized scalars responsible for coordinate dependent brane tensions. These effects, which constitute salient features of geometries with nested warping, may introduce a plethora of non-trivial modifications as far as collider signatures are concerned. Theoretically, it is worth investigating how other viable choices of $f(R)$, or other classes of higher curvature theories (eg. Einstein-Gauss-Bonnet gravity), incorporate higher order phenomena like significant back reaction and affect the stabilization scheme. Such studies would necessarily have to rely on numerical techniques due to the complexity of the model. As plausible alternatives, one can also attempt to explain the origin of the stabilizing potential from quantum and/or thermodynamic perspectives, proceeding along the lines of \cite{therm1,therm2,quant1,quant2,quant3}. Finally, it would be interesting to study the role of multiple dynamical radion fields in various cosmological contexts as well, eg. in the inflationary and bouncing settings, alongside their signature on the Cosmic Microwave Background.

\section*{Acknowledgements}
Research work of AB was partly supported by MS studentship of IACS Kolkata.

\end{document}